\newcommand{\fref}[1] {figure \ref{#1}}

\newcommand{\eref}[1] {(\ref{#1})}
\newcommand{\Eref}[1] {Eq.~(\ref{#1})}
\newcommand{\Fref}[1] {Fig.~\ref{#1}}

\newcommand{\isum}%
{\mathop{\hbox{$\displaystyle\sum\kern-13.2pt\int\kern1.5pt$}}}
 \documentclass[final,5p,times,twocolumn]{elsarticle}

\usepackage{amssymb,amsmath,mathtools}
\usepackage{graphicx,bm,epsfig}
\biboptions{sort&compress}

\journal{Computer Physics Communications}

\begin{document}

\begin{frontmatter}



\title{Solving close-coupling equations in momentum space without
  singularities for charged targets}


\author{A.~W.~Bray\corref{cor1}\fnref{1}}
 \fntext[1]{present address: Research School of Physics and Engineering, Australian National University, Canberra ACT 2601, Australia}
\cortext[cor1]{phone: +61892667747}
\ead{alexander.w.bray@graduate.curtin.edu.au}
\author{I.~B.~Abdurakhmanov}
\author{A.~S.~Kadyrov}
\author{D.~V.~Fursa}
\author{I.~Bray} 

\address{Curtin Institute for Computation and Department of Physics and Astronomy, Curtin University, GPO Box U1987, Perth, WA 6845, Australia}

\begin{abstract}
The analytical treatment of the Green's function in the convergent
close-coupling method [Bray {\em et al.} Comp. Phys. Comm. {\bf 203} 147
(2016)] has been extended to charged targets. Furthermore, we show
that this approach allows for calculation of cross sections at zero
channel energy. For neutral targets this means the electron scattering length
may be obtained from a single calculation with zero incident
energy. For charged targets the non-zero excitation cross sections at
thresholds can also be calculated by simply setting the incident
energy to the exact threshold value. These features are demonstrated by
considering electron scattering on H and He$^+$.
\end{abstract}

\end{frontmatter}


%\tableofcontents

%\include{Igor_intro}
\section{Introduction}
There has been immense progress in the field of atomic and molecular
scattering theory during the last two decades. Computational methods
such as R-matrix with pseudostates~\cite{BBS97,0953-4075-36-18-301,0953-4075-46-11-112001}, exterior complex
scaling~\cite{MBR04,B06}, time-dependent close-coupling~\cite{Pindzola07tr}, and convergent
close-coupling (CCC)~\cite{BS92,Bpr12} all set out to fully solve the
underlying Schr\"odinger equation without approximation. Collisions involving
electron, positron, or photon scattering on few-electron
atoms and ions can now be routinely calculated accurately at any
energy of interest. The CCC approach has also been recently extended
to molecular targets~\cite{ZSFB16l} and heavy projectiles like antiprotons and protons\cite{AKFB13l,AKB16,AKAFB16,AKAB16,AKB16pra}.

Generally, further progress in the field comes from extending the capability to
more complex collision systems. However, recently we found that a
novel approach to the solution of the CCC equations yields greater
utility when applied to ill-conditioned systems such as two-centre
positron-atom scattering~\cite{Bcpc16,FBKB16,CKB16}. Starting with the
electron-hydrogen S-wave model, it was shown that the Green's function
in the CCC coupled Lippmann-Schwinger equations may be treated
analytically, and thereby removing a somewhat problematic
principal-value integral~\cite{Bcpc15}. The full
implementation~\cite{Bcpc16} is applicable to electron or positron
scattering on neutral
targets. However, in the case of ionic targets there is considerable
extra complexity due to the requirement for the inclusion of projectile
bound states. Extension to ionic targets is useful in its own right,
and is also a requirement for the application of the CCC method to
photon scattering~\cite{KB96,KB01Be,Cetal00,Hetal09l}.

Another interesting consequence of the analytical approach is that the
elimination of the singularity allows the application of the method at
exact threshold energies. In the case of neutral targets this is 
useful when calculating the scattering length. However, for charged
targets the excitation cross sections are non zero at threshold, and
as we shall show, may be directly calculated.
 Atomic units are used
        throughout unless specified otherwise.
\section{Theory}
The general ideas behind the analytic treatment of the Green's
function in the Lippmann-Schinger equations have already been
discussed earlier~\cite{Bcpc15,Bcpc16}. Here we concentrate on the
extra complexity associated with charged targets. In such cases the
projectile asymptotic Hamiltonian contains the potential due to the
asymptotic  nuclear charge $Z_{\rm a}$, leading to the projectile wave
satisfying 
  \begin{equation}
    \label{eq:proj}
    \left(K+\frac{z_{\rm p}Z_{\rm a}}{r}-\varepsilon_k\right)|\bm{k}_{\rm p}\rangle=0,
  \end{equation}
where $K$ is the projectile kinetic energy operator and $\bm{k}_{\rm
  p}$ is projectile momentum.
In the case of electron scattering we have $z_{\rm p}=-1$, and for the
He$^+$ target $Z_{\rm a}=+1$. Consequently, the complete set in
projectile space includes the countably infinite number of discrete
states as well as the continuum. The resulting coupled Lippmann-Schwinger
equations~\cite{B94} take almost an identical form to the case of
neutral targets except for the requirement to also include the bound states
of the projectile of energy $\varepsilon_k$
	\begin{alignat}{3}
		\label{CCCorig}
		\nonumber
            \langle \bm{k}_f \phi_f | T_S | \phi_i \bm{k}_i\rangle =\; &\langle \bm{k}_f \phi_f | V_S | \phi_i \bm{k}_i\rangle\\
		&+ \sum_{n=1}^{N} \isum \! \mathrm{d}\bm{k} \frac{\langle \bm{k}_f \phi_f | V_S | \phi_n \bm{k}\rangle 
		\langle \bm{k} \phi_n | T_S |\phi_i \bm{k}_i\rangle}{E+i0-\epsilon_n-\varepsilon_k}.
	\end{alignat}
	Here $\bm{k}_f$ ($\bm{k}_i$) is the projectile final (initial)
        momentum,  the notation
	$i0$ is used to indicate the limit of $ix$ as positive $x\to0$
        to ensure outgoing spherical wave boundary conditions, $S$ is the total spin of the system, $V_S$ are the interaction potentials
	and $T_S$ are the required transition amplitudes. For He$^+$
        the target wavefunctions $\phi_n$ of energy $\epsilon_n$ are obtained from the Hamiltonian
        in \Eref{eq:proj} with $Z_{\rm a}$=+2.

\Eref{CCCorig} is solved by first taking the complex part of the
integral analytically to yield the equation for the $K$
matrix~\cite{BS92}
	\begin{alignat}{3}
		\label{CCCorigK}
		\nonumber
            \langle \bm{k}_f \phi_f | K_S | \phi_i \bm{k}_i\rangle =\; &\langle \bm{k}_f \phi_f | V_S | \phi_i \bm{k}_i\rangle\\
		&+ \sum_{n=1}^{N} {\cal P}\isum \! \mathrm{d}\bm{k} \frac{\langle \bm{k}_f \phi_f | V_S | \phi_n \bm{k}\rangle 
		\langle \bm{k} \phi_n | K_S |\phi_i \bm{k}_i\rangle}{E+i0-\epsilon_n-\varepsilon_k},
	\end{alignat}
where the required $T$ matrix is obtained from
\begin{equation}
  \label{eq:KT}
   \langle \bm{k}_f \phi_f | K_S | \phi_i \bm{k}_i\rangle =
 \sum_{n=1}^{N_{\rm o}}\langle \bm{k}_f \phi_f | T_S | \phi_n
 \bm{k}_n\rangle
\left(\delta_{ni}+i\pi k_n\langle \bm{k}_n \phi_n | K_S | \phi_i \bm{k}_i\rangle\right),
\end{equation}
and where $N_{\rm o}$ is the number of open states such that
$E-\epsilon_n=k_n^2/2>0$. The symbol ${\cal P}$ indicates a principal
value integral.

A partial-wave expansion is utilised to solve \Eref{CCCorigK}, with
the numerical details for the original approach given in
Ref.~\cite{BS95}. Briefly, the singularity in \Eref{CCCorigK}, whenever
$E-\epsilon_n=k_n^2/2>0$, is treated using symmetric, about $k_n$,
quadratures. This is 
problematic near excitation thresholds ($k_n\approx0$), and whenever
the system of equations is particularly ill-conditioned, such that the
addition of large values either side of the singularity causes
considerable precision loss. The latter circumstance is commonly the
case for for positron scattering 
treated within the two-centre formalism~\cite{KB02}.

An alternative approach has been proposed, which treats the integral
in the partial-wave expanded \Eref{CCCorigK} analytically~\cite{Bcpc15,Bcpc16}. For neutral
targets, following the introduction of several complete sets of states,
the integral can be isolated to
	\begin{align}
		\label{G} 
			G^L_{n}(r',r'')&=\mathcal{P}\int_0^\infty \!
                        \mathrm{d}k\, \frac{\langle
                          r'|kL\rangle\langle Lk|r''\rangle}{E-\epsilon_{n}-\frac{1}{2}k^2}\nonumber\\
				     &= \mathrm{Re}\left[-\pi k_{n}^{-1}
                                       s_L\!\left(k_{n}
                                         r_<\right)\left(c_L\!\left(k_{n} r_>\right)+i\,s_L\!\left(k_{n} r_>\right)\right)\right],
		\end{align}
	where $s_L(k_nr)=\langle r|k_nL\rangle$ and $c_L(k_nr)$ are the
        regular and irregular Riccati-Bessel
        functions respectively, with
        $k_{n}=\sqrt{2(E-\epsilon_{n})}$. Note that for closed channels
        $E-\epsilon_n<0$, and $k_n$ is purely imaginary.

The possibility of solving \eref{CCCorig} at zero incident energy, so
as to obtain the scattering length (via $\sigma_S=4\pi a_S^2$) from a
single calculation, was 
not considered previously, and we do so now. For $k_i^2/2=0$ we only
require $L=0$ for a non-zero cross section, and so \Eref{G}
becomes
	\begin{align}
		\label{G0} 
			G^0_{i}(r',r'')
				     =-\pi r_<. 
                                   \end{align}
There is no need to consider threshold excitation energies for neutral
targets because the $(kr)^{(L+1)}$ behaviour of $s_L(kr)$ for small $k$
ensures zero cross sections at threshold.

For charged targets the situation is somewhat more complicated. Now we
utilise the result used in the coupled-channel optical approach~\cite{BMM90,BM93}
	\begin{align}
		\label{Gc} 
			G^L_{n}(r',r'')&=\sum_{n'=1}^\infty
                        \frac{\langle r'|n'L\rangle\langle Ln'|r''\rangle}{E-\epsilon_{n}+\frac{1}{2}Z^2_{\rm a}/n'^2} 
+\mathcal{P}\int_0^\infty \!\! \mathrm{d}k\frac{\langle
  r'|kL\rangle\langle Lk|r''\rangle}{E-\epsilon_{n}-\frac{1}{2}k^2}\nonumber\\
				     &=\mathrm{Re}\left[- \pi
                                       k_{n}^{-1} f_L\!\left(k_{n}
                                         r_<\right)(g_L\!\left(k_{n}
                                         r_>\right) +i\,f_L(k_nr_>))\right], 
		\end{align}
where $\langle r|nL\rangle$ are the bound states, with energy $-\frac{1}{2} Z_{\rm
  a}^2/n^2$, and  $f_L(k_nr)=\langle r|k_nL\rangle$ and $g_L(k_nr)$ are the
corresponding regular and irregular Coulomb 
functions of energy $E-\epsilon_n=k_n^2/2$ (negative for closed channels), respectively. 

  \begin{figure}[th]
    \centering
    \includegraphics[width=\columnwidth]{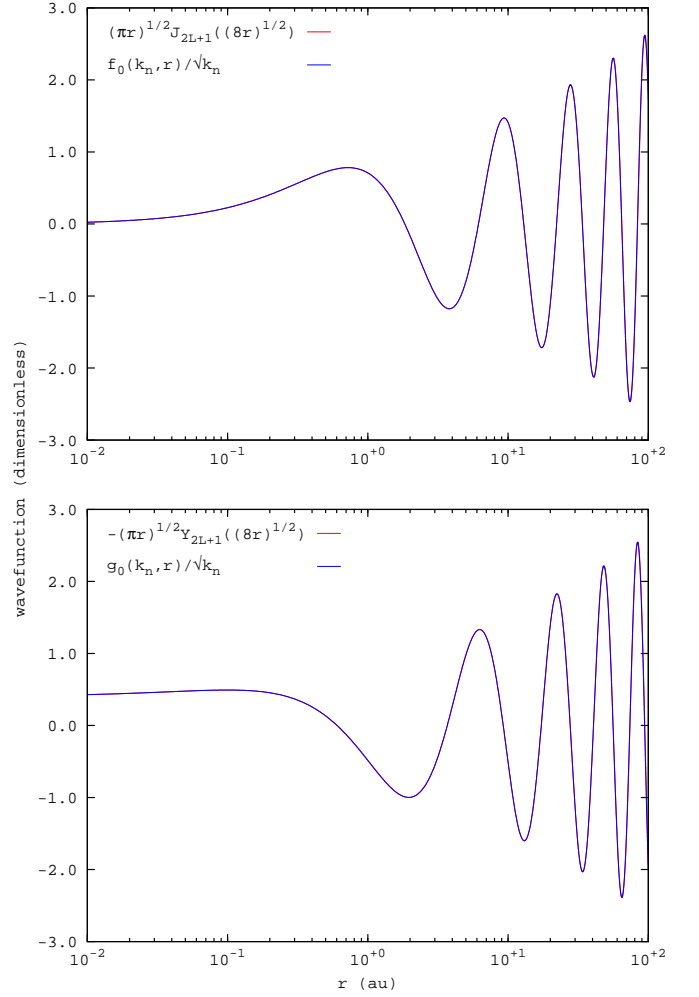}
    \caption{Regular (top) and irregular (bottom) $L=0$ Coulomb waves for zero energy
      calculated using the RHS of \Eref{eq:coul0} and the LHS with
      $k_n=0.12$. The two results are almost indistinguishable over
      the full range of $r$.}
    \label{fig:c0}
  \end{figure}

For charged targets the elastic cross section diverges as $1/k_i^2$
and so the scattering length is infinite. However, the
cross sections at excitation thresholds corresponding to $k_n=0$ are
finite for all $L$, and may be calculated in the following way. We
use
\begin{eqnarray}
  \label{eq:coul0}
  \lim_{k_n\to0}f_L(k_nr)/\sqrt{k_n} &=& \sqrt{\pi r}J_{2L+1}(\sqrt{8r}),\cr
   \lim_{k_n\to0}g_L(k_nr)/\sqrt{k_n} &=& -\sqrt{\pi r}Y_{2L+1}(\sqrt{8r}),
\end{eqnarray}
where $J$ and $Y$ are the cylindrical Bessel functions. This is
demonstrated in \Fref{fig:c0}, and allows us to eliminate the division
by $k_n$ in \Eref{Gc}.

A further problem is that the cross section is defined as
$\sigma_{fi}=k_f/k_i|T_{fi}|^2$, with $k_f=0$ at
threshold. Accordingly, we modify \Eref{eq:KT} by multiplying both
sides by $\sqrt{k_fk_i}$, which allows the removal of explicit
multiplication by $k_n$, leaving matrix elements of the form
$T_{fi}\sqrt{k_fk_i}$, which are non-zero for ions as $k_f\to0$. This
leads to a non-zero cross section for $k_f=0$ and $k_i>0$.

Having defined $G_n^L$ for all $k_n$, we first define
$N_k$ box-based 
states by solving \Eref{eq:proj} for each $L$ with a specified box size
$R_k$. For $Z_{\rm a}>0$ these will include as many bound states
($\varepsilon_k<0$) as $R_k$ allows. We then
proceed as previously~\cite{Bcpc16}, and define
	\begin{alignat}{3}
		\label{V'}
		&\langle L_f k_f l_f n_f | V'_{SJ} | n \;l \;k'' L\rangle=\int_0^\infty\!\mathrm{d}r'\int_0^\infty\;\mathrm{d}r''\sum_{k'}\nonumber\\
		&\times\langle L_f k_f l_f n_f| V_{SJ} |n\;l \;k'L\rangle \langle L\; k'|r'\rangle G^L_{nl}(r',r'')\langle r''|k''L\rangle, 
	\end{alignat}
	with the final equation to be solved being
	\begin{alignat}{3}
		\label{CCCnew}
            &\langle L_f k_f l_f n_f| K_{SJ} | n_i l_i  k_i L_i\rangle = \langle L_f k_f l_f n_f| V_{SJ} |n_i l_i  k_i L_i \rangle \nonumber\\
		&+ \sum_{l,L}\sum_{n=1}^{N}
            \sum_k\langle L_f k_f l_f n_f | V'_{SJ} | n \;l \;k\; L\rangle \langle L\; k\; l\; n | K_{SJ} |n_i l_i  k_i L_i\rangle,
	\end{alignat}
	which has no singularities. Convergence in the solution of
        \Eref{CCCnew} is obtained with increasing $N_k$ and
        $R_k$. Note that there are no substantially extra computational
        resources required in the evaluating \Eref{V'} due to the
        near-separable nature of $G_n^L(r',r'')$.

\section{Results}
We begin by considering the e-H scattering system to
demonstrate that both the numerical (nGF) and analytical (aGF)
approaches to treating the Green's function in
\Eref{CCCorig} yield the same results at all energies. Furthermore,
only the analytical approach may be applied at exactly zero projectile
energy. The zeroth partial wave suffices for our purposes. The target
states are taken to have $l_{\rm max}=2$ and $N_l=10-l$ with Laguerre
exponential parameter $\lambda_l=2$. 
An energy range varying over six
orders of magnitude is considered, though no attempt is made to make
it sufficiently dense to map out any resonance structures. 

\begin{figure}[htb]
		\centering
		\includegraphics[width=\columnwidth]{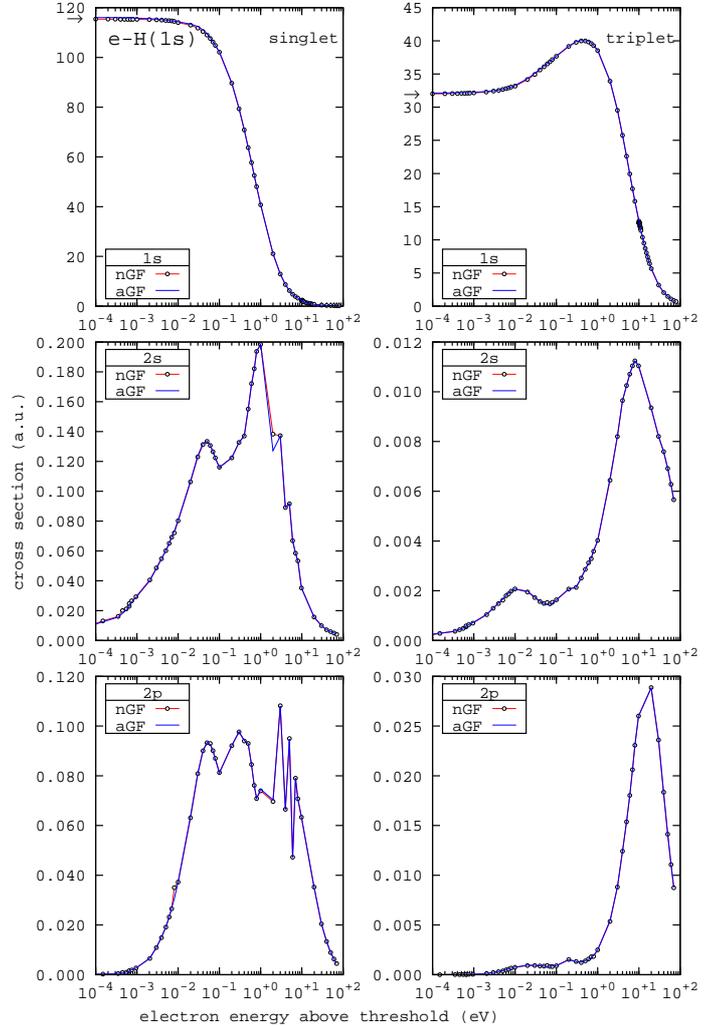}
		\caption{Electron-hydrogen scattering spin-weighted cross
                  sections for the zeroth partial wave calculated 
                  with the numerical (nGF) and analytical (aGF)
                  treatment of the Green's function in \eref{CCCorigK}. The
                  arrows indicate the $\frac{2S+1}{4}\sigma_S$ results when the projectile
                  energy is set to zero, allowing for the determination
                  of the scattering length $a_S$ via $\sigma_S=4\pi
                  a_S^2$ from a single calculation. The calculations
                  have been performed at the specified points
                  connected by straight lines to help guide the eye.
		\label{fig2}
                }
	\end{figure}

The e-H cross sections for elastic scattering and 2s, and 2p
excitation are presented in \Fref{fig2}. We see excellent agreement
between the two approaches across all energies and for all
transitions. For elastic scattering the arrows indicate the aGF
calculation performed at exactly zero energy. As required, it is
consistent with those performed at very small energies, and yields a
scattering lengths $a_{S=0}=6.0$ and $a_{S=1}=1.8$, which are
consistent with those of \citet{Schwartz61}. Note that
the cross sections are plotted against energy above threshold, which
emphasizes the near-threshold behaviour, where they start from zero.

\begin{figure}[htb]
		\centering
		\includegraphics[width=\columnwidth]{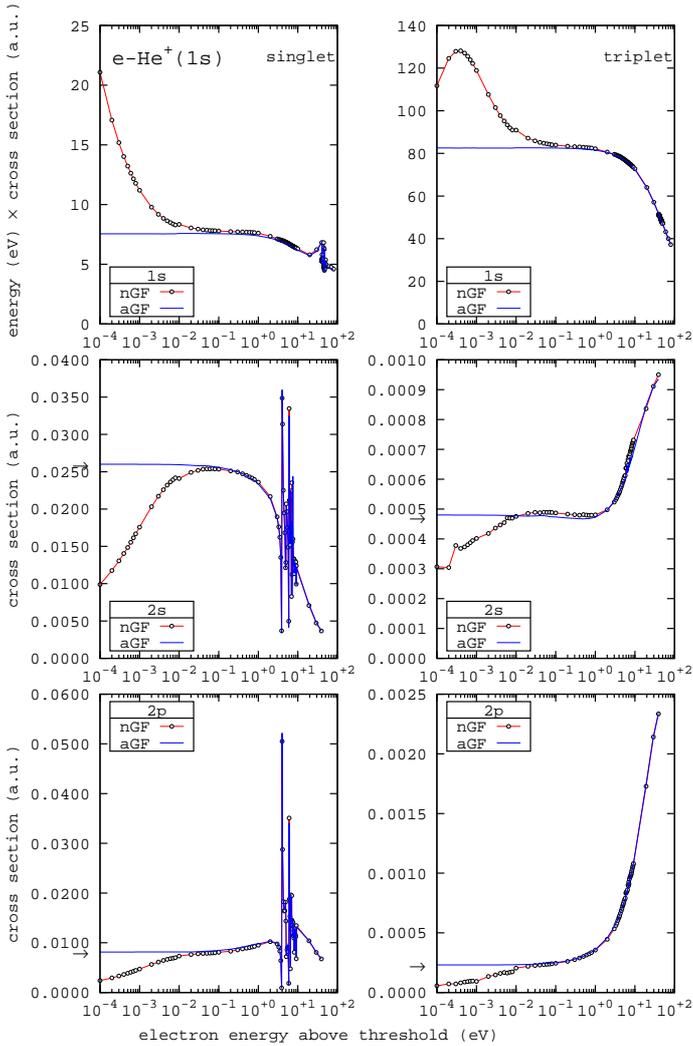}
		\caption{As for \fref{fig2} except for the He$^+$
                  target. Now the
                  arrows indicate the (non-zero) cross sections when
                  the projectile 
                  energy is set to the threshold of excitation. Note
                  that since the elastic cross section diverges as
                  $1/E_{\rm p}$, it has been multiplied by the
                  incident projectile energy $E_{\rm p}$.
		\label{fig3}
                }
	\end{figure}

We next consider e-He$^+$ scattering using the same parameters as for
the e-H system, with the cross sections presented in \Fref{fig3}. The
elastic cross sections have been multiplied by the incident energy to check the
expected divergence. It is clear that the original nGF approach
struggles to yield the correct behaviour close to thresholds in all
presented cases. While we can improve the given nGF results,
this requires going out to very large radial coordinates so
as to incorporate more bound states. In the original formulation the
approach to summation over the bound states and integration over the
continuum are entirely separate. However, in the aGF approach
increasing $R_k$ systematically affects both. Furthermore, for
excitation, the aGF approach is clearly consistent with calculations
exactly at the thresholds, as indicated by the arrows. It should be
said that the presentation deliberately emphasises the near threshold
energies. Away from these regions the agreement between the nGF and
aGF calculations is excellent.
Though not presented, we have checked that the behaviour for higher
partial waves is much the same as presented here for the zeroth partial wave.

\section{Conclusions}
The analytic approach to treating the Green's function in the coupled
Lippmann-Schwinger equations of the CCC method has been implemented
for charged targets, and at threshold energies. As a consequence it
has the same utility as the original numerical approach for electron,
positron or photon scattering. Earlier we have found that the
analytical approach yields less ill-conditined systems in the case of
two-centre positron-atom scattering~\cite{Bcpc16,FBKB16}. Here we have
demonstrated a more accurate treatment of the electron-ion collision system
in the vicinity of thresholds. 
\section*{Acknowledgements}
	This work was supported by resources provided by the Pawsey Supercomputing
	Centre with funding from the Australian Research Council,
	Australian Government and the Government of Western Australia.
ASK acknowledges partial support from the US National Science Foundation under Award No. PHY-1415656.


%% References

%\bibliography{references}

\end{document}